# Pressure-controlled interlayer magnetism in atomically thin CrI$_3$

Tingxin Li[1], Shengwei Jiang[2], Nikhil Sivadas[1], Zefang Wang[1], Yang Xu[1], Daniel Weber[3], Joshua E. Goldberger[3], Kenji Watanabe[4], Takashi Taniguchi[4], Craig J. Fennie[1], Kin Fai Mak[1,2,5]*, and Jie Shan[1,2,5]*

[1] School of Applied and Engineering Physics, Cornell University, Ithaca, NY 14853, USA
[2] Laboratory of Atomic and Solid State Physics, Cornell University, Ithaca, NY 14853, USA
[3] Department of Chemistry and Biochemistry, The Ohio State University, Columbus, OH 43210, USA
[4] National Institute for Materials Science, 1-1 Namiki, Tsukuba, 305-0044, Japan
[5] Kavli Institute at Cornell for Nanoscale Science, Ithaca, NY 14853, USA

*Correspondence to: km627@cornell.edu, jie.shan@cornell.edu
These authors contributed equally: Tingxin Li, Shengwei Jiang.

**Stacking order can significantly influence the physical properties of two-dimensional (2D) van der Waals materials[1]. The recent isolation of atomically thin magnetic materials[2-22] opens the door for control and design of magnetism via stacking order. Here we apply hydrostatic pressure up to 2 GPa to modify the stacking order in a prototype van der Waals magnetic insulator CrI$_3$. We observe an irreversible interlayer antiferromagnetic (AF) to ferromagnetic (FM) transition in atomically thin CrI$_3$ by magnetic circular dichroism and electron tunneling measurements. The effect is accompanied by a monoclinic to a rhombohedral stacking order change characterized by polarized Raman spectroscopy. Before the structural change, the interlayer AF coupling energy can be tuned up by nearly 100% by pressure. Our experiment reveals interlayer FM coupling, which is the established ground state in bulk CrI$_3$, but never observed in native exfoliated thin films. The observed correlation between the magnetic ground state and the stacking order is in good agreement with first principles calculations[23-27] and suggests a route towards nanoscale magnetic textures by moiré engineering[28].**

Intrinsic magnetism in 2D van der Waals materials has received growing attention[2-22]. Of particular interest is the thickness-dependent magnetic ground state in atomically thin CrI$_3$. In these exfoliated thin films, the magnetic moments are aligned (in the out-of-plane direction) in each layer, but anti-aligned in adjacent layers[3,12-22]. They are FM (or AF) depending on whether there is (or isn't) an uncompensated layer. The relatively weak interlayer coupling compared to the intralayer coupling allows effective ways to control the interlayer magnetism, which have led to interesting spintronics applications including voltage switching [12-14], spin filtering[16-20] and spin transistors[21]. The origin of interlayer AF coupling is, however, not well understood since interlayer FM order is the ground state in the bulk crystals. Recent *ab initio* calculations[23-27] and experiments[22,29,30] have suggested that stacking order could provide an explanation but a direct correlation between stacking order and interlayer magnetism is lacking.

In bulk CrI$_3$, the Cr atoms in each layer form a honeycomb structure, and each Cr atom is surrounded by six I atoms in an octahedral coordination (Fig. 1a). The bulk crystals undergo a structural phase transition from a monoclinic phase (space group *C2/m*) at room temperature to a rhombohedral phase (space group $R\bar{3}$) at around 210-220 K (Ref.[31,32]) with a small volume reduction. The major difference between the two phases is the stacking order while the in-plane structure remains nearly unchanged. In the monoclinic phase, each layer is displaced by a translation vector of (1/3, 0), while in the rhombohedral phase, by (1/3, -1/3) in the ABC order (Fig. 1a). The structural change can be characterized by polarized Raman spectroscopy (see Methods for details). Figure 1b is the spectra recorded at 300 K and 90 K in the back-scattering crossed-polarization configuration. The peak near 107 cm$^{-1}$ is a two-fold degenerate $E_g$ mode for the rhombohedral phase, whose energy and intensity are independent of the polarization angle (right panel, Fig. 1c). In contrast, for the monoclinic phase with lower symmetry, the mode splits into an $A_g$ and $B_g$ mode with distinct selection rules[29,32], which give rise to one unresolved peak with a four-fold polarization dependence for the peak energy (left panel, Fig. 1c).

In this study, we apply hydrostatic pressure on exfoliated CrI$_3$ thin films of 2 – 9 layers to investigate the pressure effect on the material's magnetic properties. Because of the strong structural anisotropy, hydrostatic pressure on layered materials modifies practically only the interlayer structure such as the layer separation[33,34] and stacking order. For all measurements we used samples in the form of tunnel junctions with atomically thin CrI$_3$ as a barrier and few-layer graphene as electrodes (Fig. 2a). The devices are encapsulated with hexagonal boron nitride (hBN) to protect CrI$_3$ from the environmental effect. A piston high-pressure cell compatible with electrical transport measurements was used to apply pressure up to 2 GPa. The pressure effect on the magnetic properties of CrI$_3$ was monitored by the tunnel magnetoresistance (TMR). Magnetic circular dichroism (MCD) microscopy and polarized Raman spectroscopy were also employed to characterize the magnetic state and the crystal structure of CrI$_3$, respectively, before and after applying pressure. The MCD data were taken at about 3.5 K, and the tunnel conductance data, at about 1.7 K unless otherwise specified. Since pressure can be varied only at room temperature, the samples typically went through many thermal cycles (under zero magnetic field) during the measurements. (See Methods for details on the bulk crystal growth and device fabrication, as well as the pressure cell and the MCD measurements.)

Figure 2b is an optical micrograph of an exfoliated CrI$_3$ film before being integrated into a tunnel junction. The film consists mostly of 2-layer and 5-layer regions. We applied pressure of 1.8 GPa on the sample at room temperature, cooled it down to 1.7 K with pressure, and released pressure after the sample was warmed back up to room temperature. Figure 2c shows the MCD image of the sample at 633 nm under zero field before (left) and after (right) applying pressure. Accordingly, figure 2d is the magnetic-field dependence of MCD at selected locations. The MCD probes the sample magnetization, but its magnitude in samples of different thicknesses cannot be compared directly due to the variations in their spectral response and local field factor.

Before applying pressure, the MCD signal under zero field is nearly vanishing across the entire 2-layer region, and is finite (but generally weak) in the 5-layer region. This is consistent with the interlayer-AF order: 2-layer CrI$_3$ is an antiferromagnet, and 5-layer, a ferromagnet due to an uncompensated layer. This is further seen in Fig. 2d (left panel). In addition to the FM loop centered at zero field, both 5-layer regions exhibit two spin-flip transitions: one near 0.75 T and the other near 1.7 T. These transitions correspond to spin flip at the surface layers and the interior layers of the sample, respectively. Above the second transition field, the magnetic moments in all layers are aligned. Since interior layers are absent in bilayers, there is only one spin-flip transition near 0.75 T. At these transitions (first-order phase transitions), hysteresis is clearly visible. These observations are in good agreement with the reported results[3,12-17]. The MCD image also shows the presence of inhomogeneities. In particular, the nonvanishing signal in even-layer-number regions and the significantly higher signal in certain 5-layer regions are likely from stacking faults which can introduce interlayer FM coupling as we discuss below.

Results become dramatically different after applying pressure (right panel, Fig. 2c, d). A significantly higher MCD signal at zero field emerges almost everywhere, suggesting potentially an interlayer AF-to-FM phase transition in atomically thin $CrI_3$. This is supported by the magnetic-field dependence of MCD in both 5-layer regions and the second 2-layer region (denoted by a 'red' dot in Fig. 2b): a clear hysteresis loop centered at zero field is observed. The FM phase persists to about 60 K in both 5-layer and 2-layer regions (Fig. 2e), which agrees well with the Curie temperature of the bulk crystals. The MCD result also shows that the pressure-induced AF-to-FM phase transition is not complete in many locations across the sample. For instance, a mixed FM and AF response with a visible spin-flip transition is observed over a probe area of about 1 $\mu m^2$ in the first 2-layer region ('black' dot in Fig. 2b).

Next we correlate the pressure-induced interlayer AF-to-FM transition with the crystal structure by performing polarized Raman measurements similar to that on the bulk crystals in Fig. 1b, c. Because the Raman efficiency decreases rapidly with sample thickness, we focus only on the 5-layer region ('green' in Fig. 2b). Figure 3 shows the polarization angle dependence of the Raman mode near 107 $cm^{-1}$ in the crossed-polarization configuration (See Supplementary Sect. 3 for more Raman data). Before applying pressure, the Raman response shows a clear four-fold pattern with a maximum peak shift of about 1.5 $cm^{-1}$ at 300 K (Fig. 3a). This is consistent with the monoclinic structure observed in the bulk crystals at high temperature. However, in contrary to the bulk case, the four-fold pattern remains on cooling to 10 K (Fig. 3c for 90 K), and the rhombohedral phase is not observed. This observation is consistent with a recent Raman study of $CrCl_3$ of 17 nm in thickness[29] and a second harmonic generation study of few-layer $CrI_3$ (Ref. [30]). Atomically thin crystals exfoliated at room temperature are likely kinetically trapped in the monoclinic phase at low temperature by the encapsulation layers (graphene or hBN). This is possible since the energy difference between the two stacking polytypes is small (Ref. [23-27]). After applying pressure (Fig. 3b, d), the Raman response changes to that for the rhombohedral structure: no polarization dependence can be observed for the peak energy within an uncertainty of 0.2 - 0.3 $cm^{-1}$. The response at 90 K and 300 K does not change, again supporting that the stacking order is locked in encapsulated atomically thin films. Our result therefore shows that there is a pressure-induced monoclinic-to-rhombohedral stacking order change in exfoliated thin $CrI_3$ films, and interlayer AF (FM) coupling is preferred in the monoclinic (rhombohedral) phase. This correlation between the interlayer magnetic ground state and the stacking order agrees with the recent *ab initio* calculations[23-27].

We have investigated more than ten samples, several of which consist of regions of different thicknesses. The effect of pressure and cooling on the magnetic state of these samples is summarized in Supplementary Table S1. We find that in general minimum pressure of 1.7 GPa with cooling is required to achieve a relatively thorough phase transition in bilayer $CrI_3$ and the requirements become less stringent for thicker films. Such a result can be understood from a simple energy consideration. From the bulk data, we obtain that at low temperature the rhombohedral phase has a lower energy than the monoclinic phase, $F_{R\bar{3}} - F_{c2/m} < 0$. But a structural phase transition in atomically thin samples on cooling is prevented by an energy barrier imposed by the capping layers and/or substrates. The application of pressure $P$ could facilitate such a structural phase transition by further increasing the energy difference, $P(V_{R\bar{3}} - V_{c2/m}) < 0$, since the rhombohedral structure has a smaller volume $V$. The transition is not reversible.

Finally, we perform tunneling measurements on 2D $CrI_3$ junctions to study the pressure effect on magnetism in-situ. Figure 4a shows the magnetic-field dependence of the tunnel conductance $G$ of the middle 2-layer flake (Fig. 2b) as a function of pressure. (See Supplementary Fig. S1 for the optical image of the device.) Before applying pressure, $G$ shows a jump at the spin-flip transition field $B_{sf} \approx 0.75$ T. This is the spin-filtering effect that has been recently reported in few-layer $CrI_3$ (ref [16-20]). The electron tunneling rate is higher when spins are aligned with the magnetization of each $CrI_3$ layer. The spin-filtering efficiency can be characterized by TMR (normalized conductance difference above and below $B_{sf}$ by conductance below $B_{sf}$). Again, hysteresis is clearly visible. The monotonic decrease of $G$ with increasing field is a response of the graphene electrodes. At 1 GPa, we observe three major changes: the spin-filtering efficiency decreases, $B_{sf}$ shifts to a higher value, and the overall $G$ increases. With a further increase of pressure to 1.8 GPa, the spin-filtering effect nearly disappears. After removing pressure, the behavior of the junction does not revert back to that before applying pressure. A similar trend is observed in a 4-layer tunnel junction under 0 – 1.4 GPa in Fig. 4b (See Supplementary Sections 4 for results from additional devices). There is a second $B_{sf}$ that corresponds to spin flip in the interior layers of $CrI_3$. Both transitions behave similarly as a function of pressure. After removing the pressure, in this case the spin-filtering effect is still clear: the efficiency is reduced, but the $B_{sf}$'s revert back to the values before applying pressure. We summarize the pressure dependence of $B_{sf}$ in the form of relative enhancement over the zero pressure value in Fig. 4c. It increases monotonically with pressure for samples of different thicknesses. The rate is roughly the same for all samples except that for the bilayers, which is about twice as large.

The spin-filtering efficiency is determined by fraction of the AF area in the junction. The significantly suppressed spin-filtering effect under high pressure (1.8 GPa) in the bilayer sample is consistent with a nearly complete structural phase transition. The residual effect under intermediate pressure (0.9-1.4 GPa) in the 4-layer sample can be attributed to a spatially mixed AF-FM phase in $CrI_3$. On the other hand, spin flip occurs when the Zeeman splitting energy becomes comparable to the interlayer AF exchange energy $J_\perp \sim g\mu_B m_S B_{sf}$, where $g \approx 2$ is the electron g-factor, $\mu_B$ is the Bohr magneton, and $m_S \approx \frac{3}{2}$ is the spin magnetic quantum number of $Cr^{3+}$ cations[31]. In bilayer $CrI_3$, $J_\perp$ is about 0.25 meV with $B_{sf} \approx 0.75$ T under zero pressure, and it nearly doubles under 1.5 GPa. Our first-principles calculations show that for bilayer $CrI_3$ in space group $C2/m$, $J_\perp$ remains AF, and its magnitude increases nearly linearly with decreasing layer separation by a few percent (Supplementary Sect. 6). Result in Fig. 4c is thus consistent with the picture of layer compression without a stacking order change on application of pressure. The reduction in layer separation also contributes to the observed increase of the overall conductance. But future theoretical studies are needed for a more quantitative comparison between experiment and theory and to understand the unusually large pressure effect on the interlayer exchange interaction in bilayer $CrI_3$.

In conclusion, we have experimentally demonstrated control of the interlayer magnetic order through the crystal stacking order in atomically thin van der Waals magnets by application of hydrostatic pressure. Pressure can also be employed to strengthen the interlayer exchange interaction. Our findings not only shed light on the physical origin of the thickness-dependent magnetic ground state in atomically thin $CrI_3$, but the correlation between stacking and magnetic ordering also paves the path for engineering moiré magnetism in double-layer heterostructures of van der Waals magnets[23,28].

**Competing interests**

The authors declare no competing interests

**Data availability**

The data supporting the plots within this paper and other findings of this study are available from the corresponding authors upon request.

**Methods**

**Growth of CrI$_3$ bulk crystals.** Bulk CrI$_3$ crystals were synthesized by chemical vapor transport, as described previously[35]. The elements were sealed in an evacuated ampoule (109.0 mg Cr chunks, 1 eq., Alfa Aesar, 99.999 % purity; 798.0 mg I$_2$ crystals, 1.5 eq, Alfa Aesar, 99.999 % purity). The ampoule was heated to 650 °C at the feed and 550 °C at the growth zone. After a week, the ampoule was cooled and the reaction yielded crystals with 1-2 mm edge length. CrI$_3$ produced by this method crystallized in the *C*2/*m* space group with typical lattice parameters of *a* = 6.904 Å, *b* = 11.899 Å, *c* = 7.008 Å and *β* = 108.74 ° at room temperature. In a typical sample, the crystals exhibit a Curie temperature of 61 K as well as a structural phase transition to the space group $R\bar{3}$ at 210-220 K[31].

**Device fabrication.** CrI$_3$ tunnel junctions were fabricated by the layer-by-layer dry-transfer method[36]. Atomically thin flakes of hexagonal boron nitride (hBN), graphite, and CrI$_3$ were mechanically exfoliated from bulk crystals onto silicon substrates covered by a 300-nm thermal oxide layer. The flakes were picked up in sequence by a stamp made of a thin film of polycarbonate (PC) on polydimethylsiloxane (PDMS). The entire stack was then released onto a substrate with prepatterned titanium/gold (Ti/Au 5 nm/35 nm) electrodes. The residual PC on the device surface was dissolved in chloroform before measurements. Both the exfoliation and the transfer processes were performed in a nitrogen-filled glovebox to avoid degradation of CrI$_3$. The thickness of atomically thin CrI$_3$ samples was initially estimated from their optical reflectance contrast, and later verified by MCD or atomic force microscopy measurements.

**High-pressure experiments.** A piston high-pressure cell compatible with electrical transport at low temperature (easyCell 30, Almax easyLab) was employed. The cylindrical cell has an inner bore diameter of about 3.9 mm. Hydrostatic pressure is applied on the samples through pressure transmission medium, pentane and iso-pentane mixture (1:1), which freezes around 160 K. The typical range of pressure that can be accessed at low temperature is 1 - 2 GPa.

Pressure was applied (or changed) at room temperature. The cell was then cooled to low temperature for measurements. Two types of manometers were used for pressure calibration. At room temperature, pressure was calibrated by measuring the resistance of a manganin wire (~ 10 cm long) wound into a coil[37]. At low temperature, pressure was calibrated by measuring the superconducting transition temperature of a tin wire[37]. The pressure is generally about 0.1 - 0.3 GPa lower at low temperature than at room temperature due to the freezing of the pressure transmission medium.

**MCD microscopy**. Magnetic circular dichroism (MCD) measurements were performed in an attoDry1000 cryostat with a base temperature of 3.5 K. For non-imaging measurements, a HeNe laser at 633 nm was used. The laser beam was coupled into and out of the cryostat using free-space optics. It was focused onto the sample by a cold objective. The beam size on the sample was about 1 µm$^2$, and the power, no more than 10 µW. The incident beam was modulated between left and right circular polarization by a photoelastic modulator at 50.1 kHz. The reflected beam was collected by the same objective and detected by a photodiode. The MCD signal was determined as a ratio of the ac component at 50.1 kHz (measured by the lock-in amplifier) and the dc component (measured by a digital multimeter) of the reflected light intensity.

In MCD imaging, a broad-field illumination and a nitrogen-cooled charge-coupled device (CCD) were used. The light source was selected from an incoherent white light using a band pass filter centered at 632 nm. A linear polarizer and a quarter wave plate were used to generate circularly polarized light. The MCD image was calculated as the normalized difference between the left and right circularly polarized light reflection by the total reflection. The spatial resolution is diffraction limited (~ 500 nm).

**Polarized Raman spectroscopy.** Raman spectroscopy was performed using a home-built microscope in the back-scattering geometry. A solid-state laser at 532 nm was employed as the excitation source. The laser beam was focused onto the $CrI_3$ samples along the c-axis by a 40X objective. The scattered light was collected by the same objective and detected by a spectrometer with a 1800-grooves/mm diffraction grating and a nitrogen-cooled CCD. The polarized Raman modality was used to characterize the crystal structure. To this end, the laser beam passed through a linear polarizer and a half-wave plate. The latter varies the polarization angle in the crystal a-b plane. The scattered light passed through the same half-wave plate, and a second linear polarizer, which selects the component either parallel or perpendicular to the incident beam polarization (referred to as the parallel- and crossed-polarization configuration, respectively).

The polarized Raman spectroscopy in crystals of both the $R\bar{3}$ and $C2/m$ groups has been analyzed previously[29]. In particular, for the doubly degenerate $^1E_g$ and $^2E_g$ modes (in the rhombohedral phase) and the non-degenerate $A_g$ and $B_g$ modes (in the monoclinic phase) that have been examined in this study, the Raman tensors are derived as

$$^1E_g = \begin{pmatrix} c & d & e \\ d & -c & f \\ e & f & 0 \end{pmatrix}, \quad ^2E_g = \begin{pmatrix} d & -c & -f \\ -c & -d & e \\ -f & e & 0 \end{pmatrix}, \quad A_g = \begin{pmatrix} a & 0 & d \\ 0 & c & 0 \\ d & 0 & b \end{pmatrix}, \quad B_g = \begin{pmatrix} 0 & e & 0 \\ e & 0 & f \\ 0 & f & 0 \end{pmatrix}.$$

In the crossed-polarization configuration, the Raman intensity for the two $E_g$ modes is given as $I(\,^1E_g)_{cross} = |c\sin(2\theta) - d\cos(2\theta)|^2$ and $I(\,^2E_g)_{cross} = |c\cos(2\theta) + d\sin(2\theta)|^2$, where $\theta$ is the polarization angle with respect to the a-axis. The total intensity of the two degenerate modes ($c^2 + d^2$) is polarization independent. Similar result can be derived for the parallel-polarization configuration.

The result is different for the non-degenerate $A_g$ and $B_g$ modes in the monoclinic phase. The Raman intensity of each mode shows a four-fold polarization dependence: $I(A_g)_{cross} = a^2\sin^2(2\theta)$, $I(B_g)_{cross} = e^2\cos^2(2\theta)$, but they are off by $\pi/4$. Here we have assumed $a \sim -c$, since the symmetry of $A_g + B_g$ should match with the symmetry of $^1E_g + \,^2E_g$ (Ref. 29). The total intensity of these two spectrally unresolved modes shows a four-fold polarization dependence for the peak energy. Similar result can be derived for the parallel-polarization configuration.

**Figure captions**

**Fig. 1 | Crystal structure of CrI$_3$. a,** Top: atomic structure of monolayer CrI$_3$. Cr atoms (blue balls) form a honeycomb lattice structure. Each Cr atom is surrounded by six I atoms (purple balls) in an octahedral coordination. Middle: top view of stacking order in the monoclinic (left) and rhombohedral phase (right) for three layers. The Cr network shifts in the order of blue, red and black. Bottom: side view of the corresponding stacking order for two layers. a, b, and c are the crystallographic axes. **b,** Polarized Raman spectrum of a bulk CrI$_3$ crystal at 300 K (upper) and 90 K (lower) in the crossed-polarization configuration. **c,** Polarization angle dependence of the spectrum inside the dashed box in **b**. The peak energy of the mode around 107 cm$^{-1}$ shows a four-fold dependence on the polarization angle in the monoclinic phase (left), and no dependence in the rhombohedral phase (right).

**Fig. 2 | Pressure-induced interlayer AF-to-FM transition in atomically thin CrI$_3$. a,** Schematic side view of CrI$_3$ tunnel junctions employed in this study. Atomically thin CrI$_3$ serves as a tunnel barrier, and few-layer graphene (G) (connected to Au contacts) is used as tunnel electrodes. The entire device is encapsulated with hBN. **b,** Optical micrograph of a thin flake of CrI$_3$ before being integrated into a tunnel junction. The flake has regions of different thicknesses identified by dashed lines of different colors. The scale bar is 5 $\mu$m. **c,** MCD image of the CrI$_3$ flake before (left) and after (right) applying pressure of 1.8 GPa. The data were recorded at 3.5 K under zero magnetic field. **d,** Magnetic-field dependence of MCD at 3.5 K for two 2-layer and two 5-layer regions before (left) and after (right) applying pressure. The color of the lines matches the color of the dots identified on the image in **b**. **e**, MCD vs. field at varying temperature for a 5-layer (left) and a 2-layer region (right) reveals that the pressure-induced ferromagnetism persists to ~ 60 K.

**Fig. 3 | Pressure-induced structural phase transition in atomically thin CrI$_3$.** Polarization angle dependence of Raman spectrum in the crossed-polarization configuration for a 5-layer CrI$_3$ sample before (**a**, **c**) and after (**b, d**) applying pressure of 1.8 GPa; at 300 K (**a, b**) and at 90 K (**c, d**). The change from a four-fold dependence to a no-dependence for the peak energy of the mode around 107 cm$^{-1}$ is indicative of a monoclinic-to-rhombohedral phase change. No structural phase change is observed on cooling in the absence of pressure regardless of the initial structural phase.

**Fig. 4 | Spin-filtering effect in atomically thin CrI$_3$ as a function of pressure. a, b,** Tunnel conductance $G$ versus applied magnetic field $B$ of a bilayer CrI$_3$ tunnel junction at 1.7 K under varying pressure in the order of 0, 1, 1.8 and 0 GPa. **b,** Same measurement of a 4-layer CrI$_3$ tunnel junction under varying pressure in the order of 0, 0.9, 1.4, and 0 GPa. The bias voltage for all measurements was 10 mV. **c,** Enhancement of the spin-flip transition field $\frac{B_{sf}(P)-B_{sf}(0)}{B_{sf}(0)}$ as a function of pressure $P$ from 6 different devices. The dashed straight lines are a guide to the eye.

The spin-flip transition field for each field sweep direction was estimated from the field corresponding to the steepest change of *G*. Its uncertainty was estimated from the width of the rise/fall of G with field in the transition region. $B_{sf}$ in **c** was calculated as the average for the two sweep directions.

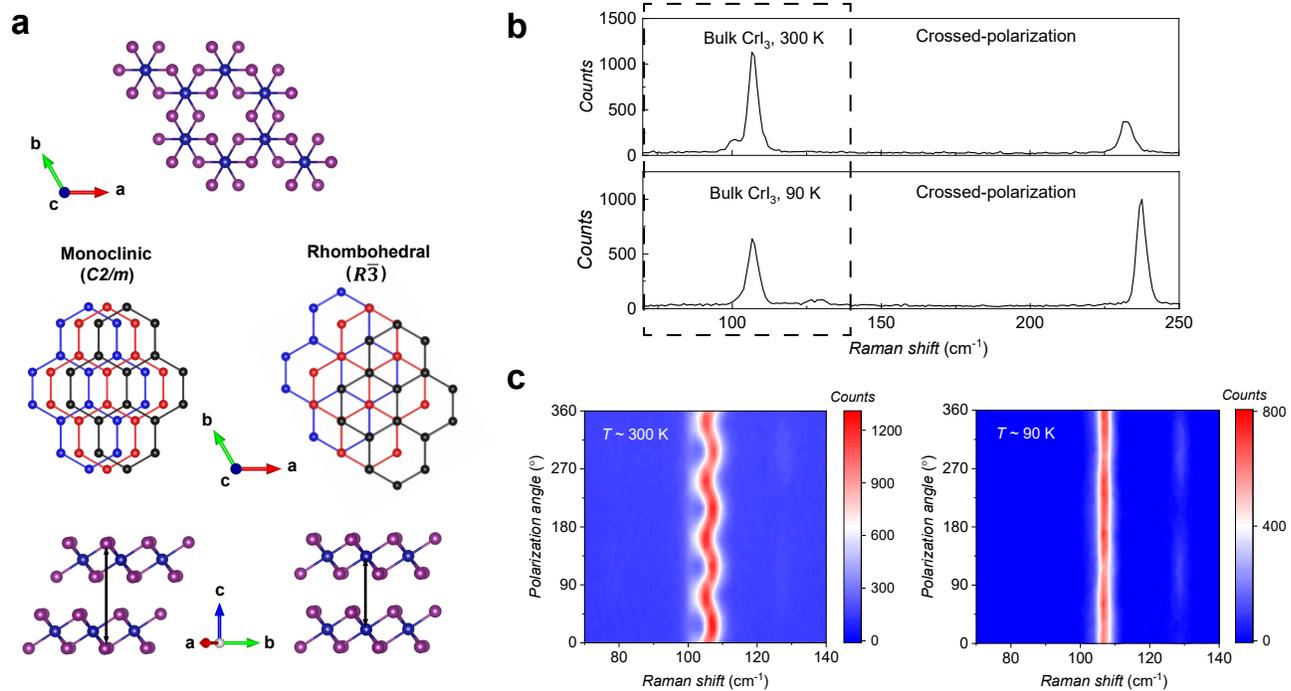

**Figure 1**

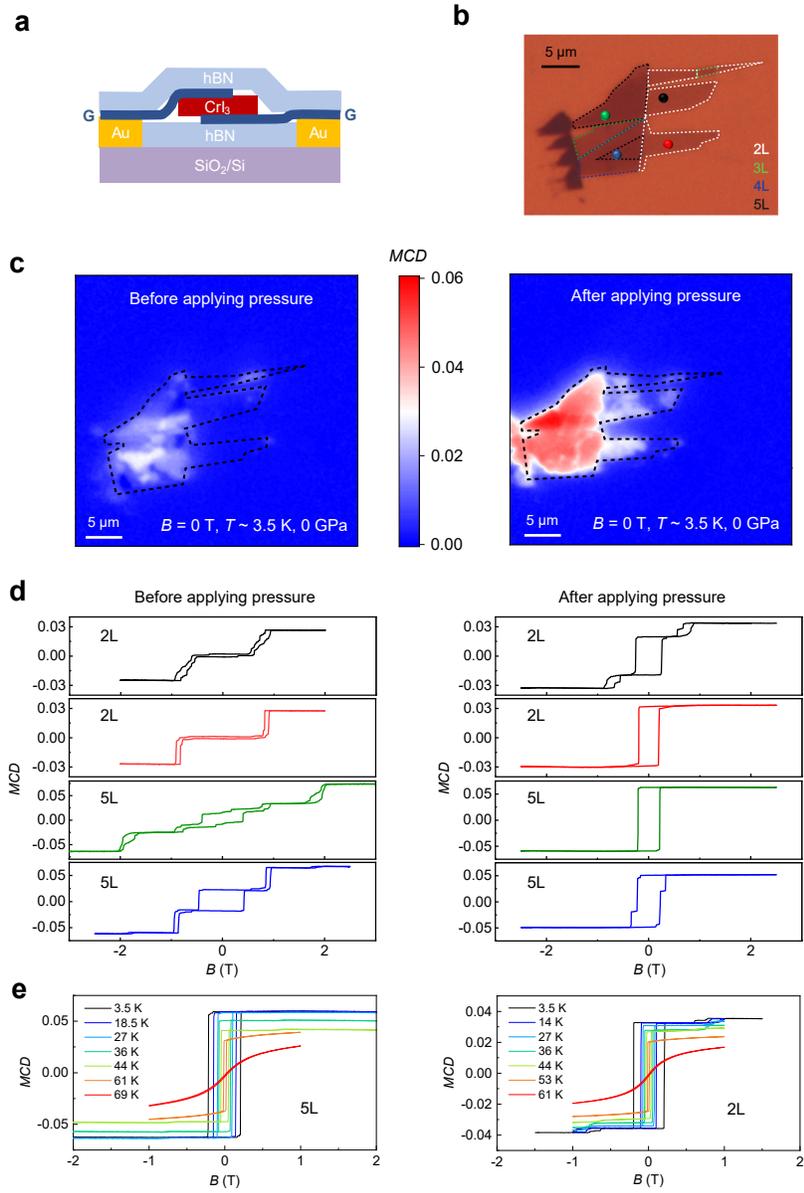

**Figure 2**

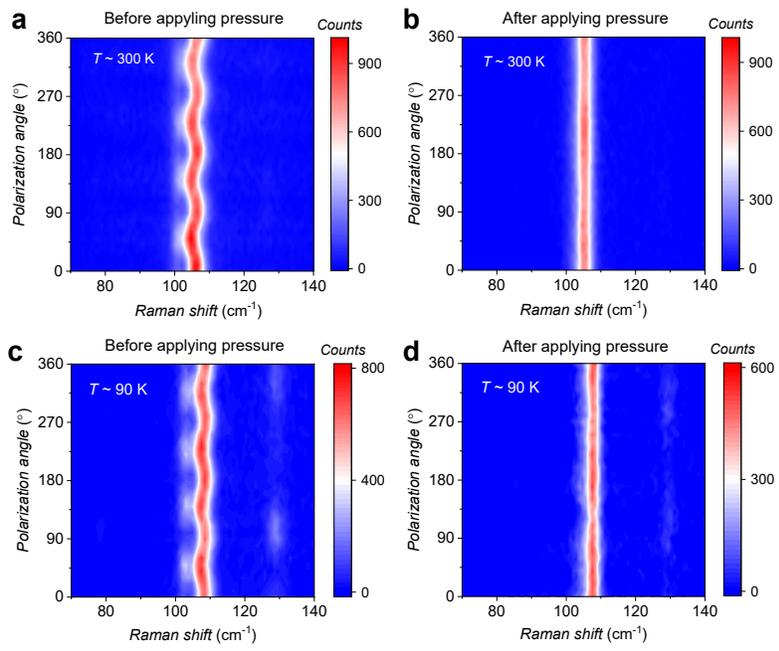

**Figure 3**

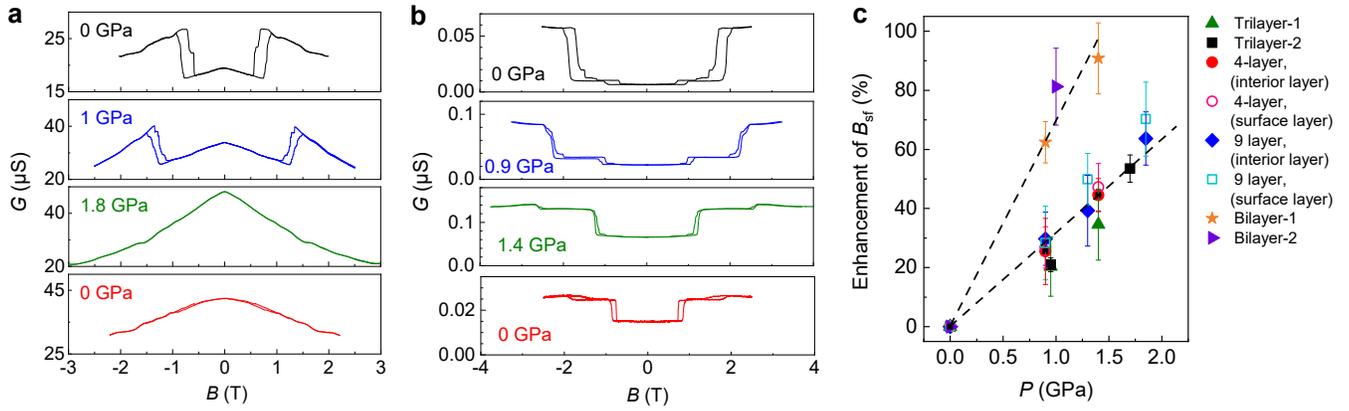

**Figure 4**